\documentclass[runningheads]{llncs}
\usepackage{graphicx}
\usepackage{epigraph}
\usepackage{tikz}
\usepackage{array}
\usepackage{appendix}
\usepackage[misc]{ifsym}

\makeatletter
\newcommand{\@chapapp}{\relax}
\makeatother

\begin{document}
	\title{Is Privacy Controllable?}
	\author{Yefim Shulman \inst{} \textsuperscript{(\Letter)} \orcidID{0000-0002-3163-9726} \and Joachim Meyer \inst{} \orcidID{0000-0002-1801-9987}}
	\authorrunning{Y. Shulman and J. Meyer}
	\institute{Tel Aviv University, Tel Aviv 6997801, Israel \\
	\email{\{efimshulman@mail.,jmeyer@\}tau.ac.il}}
	\maketitle
	
	\begin{abstract}
		One of the major views of privacy associates privacy with the control over information. This gives rise to the question how controllable privacy actually is. In this paper, we adapt certain formal methods of control theory and investigate the implications of a control theoretic analysis of privacy.
		We look at how control and feedback mechanisms have been studied in the privacy literature. Relying on the control theoretic framework, we develop a simplistic conceptual control model of privacy, formulate privacy controllability issues and suggest directions for possible research.
		
		\keywords{Privacy \and Feedback \and Information Disclosure \and Human Control \and Closed-Loop Control \and Feedback Control}
	\end{abstract}
	
	%\setlength{\epigraphrule}{0pt}
	%\setlength{\epigraphwidth}{184pt}
	%\epigraph{[...] we know nothing of our own behavior but the feedback effects of our own outputs.\\ To behave is to control perception.}{William T. Powers, \\ \textit{Behavior: The control of\\ perception (1973)}}
	
	\section{Introduction}
		Casually used in colloquial conversations, the term ``privacy'' in all its complexity and prominence appears in philosophical, legal, political, scientific and technological discussions. Although there exist numerous definitions,  Solove \cite{Solove2006taxonomy} states in his \textit{A Taxonomy of Privacy} that ``Privacy is a concept in disarray'', which ``suffers from an embarrassment of meanings'' (p. 477). Incidentally, the situation has hardly improved ever since. 
		
		While scholars struggle with privacy definitions, the general public (the users) struggle with their online privacy settings, as has been abundantly demonstrated in the security and privacy literature. The seeming futility of information control and the lack of functional transparency lead people to feel helpless.\footnote{See \cite{turow2015tradeoff} for a case of American consumers being resigned to giving up their data in exchange for commercial offers, rather than engaging in cost-benefit analyses.} 
		In spite of the diversity of approaches, many discussions of privacy tie it to some form of control (control of access to information, use of information, distribution of information, etc.). We take these terms literally.
						
		If mainstream privacy research embraces the understanding of privacy as control, is there a way to analyse the controllability of privacy? Can we borrow from formal methods of control theory to broaden our understanding of privacy issues, at least when it is appropriate to define privacy as control over information?
	
		This paper is organised in the following way. Section 2 of this paper provides the scope of the problem. Here we discuss the extent of privacy determined by control and the reach of control theory.  
		
		In Section 3 of this paper we apply a control theoretic framework for a conceptual analysis of privacy as control over information.
		
		Section 4 presents a discussion of the applicability, limitations and relevance of our analysis to contemporary research in the privacy literature.
				
		This paper is a first attempt to analyze privacy within the framework of control theory. We map control theory onto privacy, manifested as control over personal information from a user's perspective. Our analysis looks at privacy on a micro level, dealing with the topic in the meaning with which it is used in social and computer science discussions (as opposed to legal, political and philosophical interpretations). Starting from Section 3, we use the term ``privacy'' interchangeably and as a shorthand for ``personal information'' and its disclosure. Our analysis can serve as a conceptual framework for discussions of privacy and its implications in different contexts.
				   		
	\section{Privacy as Control over Information, Control as a Theory}
		
		Privacy is a permeating concept, which has no generally accepted definition throughout all disciplines. Privacy definitions\footnote{In this paper, we are not concerned with formal definitions of privacy used in cryptography and privacy-enhancing technologies (i.e., differential privacy \cite{dwork2008differential}, \textit{l}-diversity and (\textit{n,t})-closeness \cite{li2010closeness}, etc.).} are often formulated through descriptions of the features and properties of privacy and even by writing off constructs which are not privacy \cite{Smith2011IPR}.  
		Incidentally, a bibliometric analysis of computer and information ethics literature has revealed privacy as one of three major concepts in that field \cite{Heersmink2011}. It must be noted, however, that the authors make an unsubstantiated claim about differences between the American and the European approaches to privacy, based on their clustering results, where ``data protection'' fell into the ethics, rather than the privacy cluster.\footnote{The observed effect could be an artefact of their literature sample (which was not focused on- and, thus, might not be representative of privacy research), and (or) sampling method (picking selected journals in computer and information ethics without attending to the geographical and authorship scope of those journals).} 
		
		In ontological attempts to determine what privacy is, scholars often arrive at the same conclusion: general privacy is contextual. It may be internalised through different conceptualisations by different individuals \cite{Smith2011IPR}.
		
		Additional peculiarities of the concept of privacy come to light, when one is reminded that privacy and the underlying notions may be relative. For example, they may not have simultaneously direct and corresponding translations into other languages. Smith et al. (\cite{Smith2011IPR}, p. 996) write: ``Privacy corresponds to the desire of a person to control the disclosure of personal information [...]'' while ``[...] confidentiality corresponds to the controlled release of personal information to an information custodian under an agreement that limits the extent and conditions under which that information may be used or released further''. Nevertheless, the term ``privacy policy'' is conventionally used in English. Simultaneously, in any software application or website in Russian the same document is referred to as ``policy of confidentiality'' (literal translation), when it contains specifications of data processing, data protection measures and personal information collection.\footnote{In reality, ``privacy'' is directly translated as ``privateness'', while the latter corresponds to a ``degree of inviolability of private life'', whereas, in fact, ``privacy'' corresponds to several control-, protection- or jurisprudence-related terms in the Russian language (confidentiality being one).}
		
		We can, however, resort to some general conceptions and phenomena observed in the privacy literature. Thus, major approaches to define privacy in philosophical, legal and scientific writings include: privacy as control, privacy as a right, and privacy as an economic good. By and large, the meaning of privacy is attributed arguably to control over information, restriction of access, human dignity, social relationship and intimacy (\cite{decew2006privacy}, \cite{moore2008defining}, \cite{Smith2011IPR}).
		
		Privacy as control is a prominent and distinctive approach in philosophical and legal thought, and most definitions include features and properties, which are associated with the term ``control''. In fact, major theoreticians of privacy, including Warren and Brandeis \cite{warren1890right}, Fried \cite{fried1970anatomy}, and Parent \cite{parent1983pml} refer to privacy as some form of control over information. 
		
		Alan Westin defined privacy as ``the claim of individuals, groups, or institutions to determine for themselves when, how, and to what extent information about them is communicated to others'' (\cite{westin1967privacy}, p. 7).
		
		Joseph Kupfer argues that ``by providing control over information about and access to ourselves, privacy enables us to define ourselves socially in terms of intimate relationship'' (\cite{kupfer1987pasc}, p. 86).
		
		Adam Moore derives the following definition: ``A right to privacy is a right to control access to and uses of- places, bodies, and personal information'' (\cite{moore2008defining}, p. 421).
		
		Perhaps the view of privacy as control over information is so widespread, because it resonates more easily (enables operationality) with research in information systems, behavioral and cognitive psychology, and marketing management.
		
		The term ``control'' has a specific meaning in engineering, where it is used within the framework of control theory (see \cite{astrom2010feedback}, \cite{doyle2013feedback} and \cite{luenberger1979introduction} for the theoretical framework and applications). Control theory is the basis for engineering models of control of systems and processes, including those that involve a human in the control loop (see \cite{jagacinski2003control} and \cite{sheridan1974man} for applications to human performance). 
		
		Control theory has been successfully applied to the modelling of manual control over a physical system in human factors research. Concepts of control theory have been borrowed by, and have been productively adjusted to the field of social psychology \cite{MansellWarren2015TOaF}. Optimal control models in economics belong to a family of optimal control strategies of control theory. The use of computational models of behavior, including control theory, is advocated by psychology scholars in management and organisation science \cite{vancouver2012modeling}.
		Control theory has found its way into life sciences \cite{Carver2015CPPMAD}, as an alternative form of Powers' perceptual control theory (PCT)\footnote{Originates in \cite{Powers1960GFTHB}.}, when the goal of a dynamic system resides within the system itself (see \cite{Carey2014BMBNFC} for a survey of biological, neurobiological and psychological implementations of negative feedback loops).  
		
		It does not seem a stretch to assume that people get some form of feedback on their behavior, including privacy-related actions. The information people receive about outcomes of their actions may alter their behavior, which aims to reach some comfort zone, i.e., a certain level of physical, mental or emotional well-being. Of course, people may not always be able to associate the feedback with the action (and cause with effect for that matter), a growing concern for privacy researchers and data protection professionals. With recent developments, it also becomes a concern for the general public.
		
		Control theory is instrumental and productive when it is applied to phenomena where feedback plays some role. In this conceptual paper we ask what could be the implications from analyzing privacy as control in the framework of control theory?
		
		Section 3 presents our attempt to tackle this question.
			
	\section{Control Theoretic Analysis of Privacy}
		
		Control theory distinguishes between \textit{open-loop} and \textit{closed-loop (feedback) control}. In an open-loop control system, some input is fed into the system, and a process runs its course, without further interventions in the process. In closed-loop (or feedback) control the output of the process is measured, and some information about the output is provided as feedback, serving to minimize the difference between a desired state and an existing state.
		
		In the context of privacy, our system consists of:
		\begin{itemize}
			\item a person (the controller) who performs some actions (e.g., permits an app to access information about location or contacts, or posts some information on a social network); 
			\item some process that runs, depending partly on the person's actions;
			\item the controlled output, which is the disclosure of information about the person or its use; 
			\item and the evaluation of the level of disclosure of personal information\footnote{From this point on, for the sake of convenience, we may use the term ``privacy'' as a shorthand for ``personal information'' and its disclosure.}. 
		\end{itemize}
	
		Any part of the process may be affected by external factors (the environment) that may introduce noise, or disturbances.
		
\begin{figure}
			\centering
			\includegraphics[width=\textwidth]{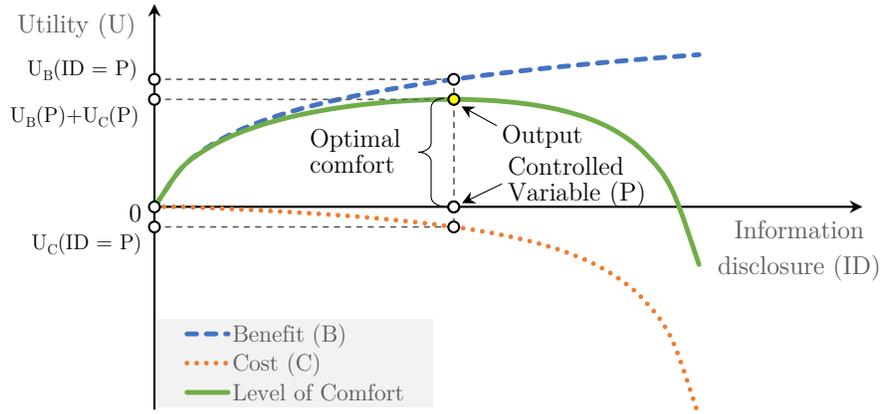}
			\caption{Privacy comfort as utility based on costs and benefits curves} \label{fig1}
		\end{figure}

		A control theoretic analysis of the user actions assumes that the output (i.e., the information disclosure) has some value that can be compared to a desired value (e.g., expressed through some personally comfortable level of disclosure). We can assume that information disclosure has some benefits (financial, emotional, social, etc.) and some possible costs. The overall outcome is the sum (or other combination) of the benefits and the costs. The exact functions by which the benefits and the costs change as more information is revealed, depend, of course, on the person, the information, the party receiving access to the information, and the specific context, in which the information is revealed. We depict a demonstration of the behavior of the controlled and output variables in Fig.~\ref{fig1}. For the sake of the demonstration, we assume that benefits increase monotonously with diminishing marginal returns, and that costs increase exponentially as the amount of revealed information increases.
		
		The change in the $Information\ Disclosure\ (ID)$ leads to desired and undesired consequences for the controller, i.e., benefits and costs, respectively. Estimating the difference between costs and benefits for each level of $(ID=P)$ while exercising control over $P$, along the abscissa axis, we seek to maximize the $Level\ of\ Comfort$ from disclosure. The character of control, as well as the optimality criterion will differ, based on the shapes of the benefits and costs functions, as defined by individual and momentary factors.
		
\begin{figure}
			\centering
			\includegraphics[width=6cm]{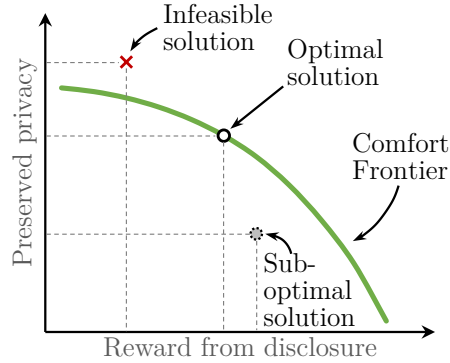}
			\caption{Privacy comfort as Pareto-optimal frontier}
			\label{fig2}
		\end{figure}

		We may assume a more complex scenario, where the control variable $P$ is multidimensional: i.e., it may contain multiple types and corresponding amounts of disclosed information $(p_{[type,amount]}\in P)$. The controller wants to maximize both pleasurable effects and privacy. The optimal comfort level can be reached when there is no possibility to improve either of the two outcomes, while maintaining the same value for the other, mapping an optimal $Output$ as a Pareto frontier, as we show in Fig.~\ref{fig2}.
		
		The Pareto frontier represents the $Output$ space, while the area under the curve contains suboptimal solutions that can be improved. The area above the frontier contains infeasible solutions, due to existing constraints on the amount of privacy preserved and the benefits gained for each level of personal information disclosure.
		
		For the person it is desirable to be on the Pareto frontier. The person has to consider two important questions: (1) Am I on the frontier? If not, what can I do to get there? (2) Where on the frontier do I prefer to be? 
		
\begin{figure}
			\centering
			\includegraphics[width=\textwidth]{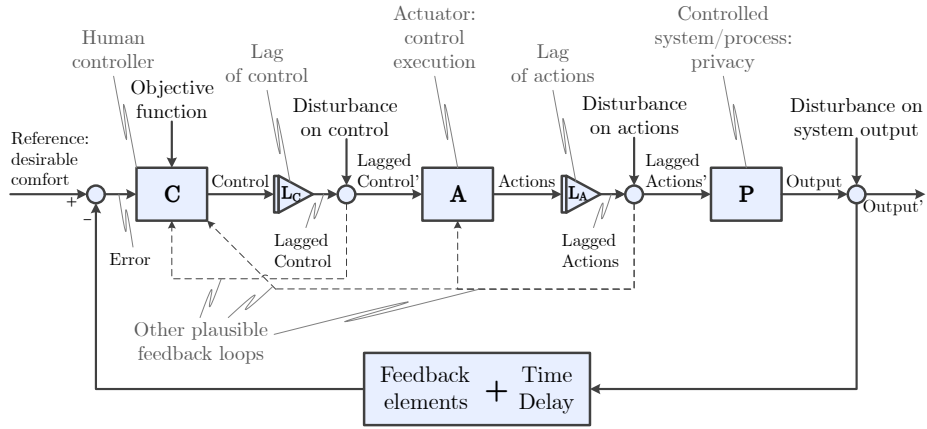}
			\caption{A block diagram representation of the privacy control model}	\label{fig3}
		\end{figure}

		Fig. \ref{fig3} contains a depiction of the proposed privacy control model in the form of a block diagram -- a widely used way to depict dynamic systems in control theory\footnote{Both in dynamic systems (e.g., \cite{astrom2010feedback}, \cite{doyle2013feedback} and \cite{luenberger1979introduction}) and human factors (e.g., \cite{jagacinski2003control} and \cite{sheridan1974man}) block diagrams are used for concise depictions of systems.}. 
		
		Each box in the block diagram is a separate subsystem. A privacy level $P$ is the amount of disclosed information, which is a controlled variable in our conceptual model. The $Output$\footnote{Or $Output'$, if any disturbance is introduced into the system after a certain $Output$ is achieved.} of our control model is the utility variable, representing the level of comfort, given the level of personal disclosure.
		
		A human controller $C$ is a person, performing actions and seeking to achieve some comfortable level of personal information disclosure. The human controller $C$ adjusts the privacy level $P$ using an actuator $A$, which is some form of a tool, system or service through which privacy-related actions are taken (e.g., settings adjustment, information sharing, etc.).
		
		Arrows represent signals flowing between the elements of the system. Arrows going into a certain element are input signals for this element, and arrows going out of an element are output signals of this element. An input signal to the whole system is $Reference$, which is some comfortable level of personal information disclosure that a person desires to achieve. An output signal of the whole system is $Output$ or $Output'$ described above. The circular blocks are ``comparators'' that sum up inflowing input signals, producing output(s). 
		
		Each triangular block represents a lag of output (that is an effect when the output of the action or process is not proportional to the input). Intermediate outputs may be non-linear, due to disturbances from the environment and the properties of the medium. The order of the lag is undefined, and the symbol is used for representation. 
	
		A time lag (delay) may also be present throughout the system. The most important time delay appears with the feedback loop (shown explicitly in Fig.~\ref{fig3}). The feedback loop includes both the information on the reached comfort value ($Output$ or $Output'$) and the time delay until the human controller receives the feedback. The time delay is a varying quantity for each individual at each point in time, which limits its predictability. 
	
		We approach the conceptual privacy control model through different topics from the systems control literature, and we reveal multiple controllability issues from the standpoint of the individual, summarized in Table~\ref{tab1}.  
		
		\begin{table}
		\caption{Privacy controllability issues}\label{tab1}
		\renewcommand{\arraystretch}{1.5}
		\renewcommand{\tabcolsep}{0.13cm}
		\begin{tabular}{>{\raggedright}p{2.2cm}|p{7.5cm}|>{\raggedright\arraybackslash}p{1.85cm}@{}}
			\hline
			\textbf{Issue} & \textbf{Description} & \textbf{Element} \\
			\hline
			Feedback time delay & The consequences of actions arrive at an uncertain time, and they may be not attributed to the actions. & Feedback loop \\ 
			Physical feedback lag & The consequences of actions may arrive non-linearly and are prone to alterations within feedback elements. & Feedback loop \\
			Multiple feedback loops & Each element of the control system (Fig.~\ref{fig3}) may have its own feedback loop(s). & Model \\
			Complexity & Elements of the system may constitute control subsystems with all the corresponding issues. & Model \\
			Order of control & Intermediate signals of controls and actions may have different non-linear profiles and may require learning from the human controller. & Forward control path \\
			Multiple physical lags & Multiple linear and non-linear relations exist between the elements of the system. & Outputs, signals \\
			Momentariness and individual differences & The privacy control model may have to be non-stationary, as privacy behavior and preferences may vary over time	for different individuals. & Concept, assumptions \\
			Linearity and time-invariance & Humans and privacy are perhaps non-linear time-variant systems: the output is not proportional to the input; %$ y(\alpha x_{i} + \beta x_{j})\neq \alpha y(x_{i}) +\beta y(x_{j})$;
			and at different points in time, the system output may differ for the same system input%: $ y(t)=f(x(t),t)\neq f(x(t)) $
			. & Concept, assumptions \\
			\hline
		\end{tabular}
		\end{table}
		These issues preclude us from asserting that humans and their personal information make up controllable systems on their own. Yet, this is not a reason for despair. It only shows that a one-to-one straightforward mapping of a control theoretic framework onto personal information disclosure cannot immediately produce beneficial results. We discuss possible implications and contributions of control theory to privacy in the following Section 4, alongside the discussion on the relevance and current standings of the notions of ``control'' and ``feedback'' in the privacy literature.
		
		In Section 4 we also proceed to discuss the contribution, applicability and limitations of our model. We further investigate the existing empirical privacy research to better understand how our conceptual analysis fares with the observed reality.  
	
	\section{Discussion}
		In Section 4.1 we present an overview of, and discussion on how privacy research has handled the notions of ``control'' and ``feedback'', and what benefits and contributions the control theoretic analysis can potentially bring. Section 4.2 describes potential research directions, driven by the control theoretic approach and clarifies the scope and limitations of this paper.
		
	\subsection{Control, Feedback and Privacy Research}
		The effects and implications of providing users with control (and the feeling of control) over their personal information and its use have been abundantly studied. Control over personal information constitutes a whole dimension of privacy concerns for users (\cite{Kitkowska2018}, \cite{Malhotra:2004:IUIPC}). Interestingly, perceived control over information does not seem to impact the level of related privacy concerns, whenever this control is perceived to be low \cite{KOWALEWSKI2015}. That result is in line with findings on feeling resigned regarding one's privacy (\cite{turow2015tradeoff}, discussed in Section 2), and that a perceived higher level of control may increase the willingness to disclose personal information \cite{brandimarte2013controlparadox}. On a more narrow approach, it has been shown that an incremental increase in controllability over information collection may make users more tolerant towards tailored online advertisement \cite{Chanchary2015UPShAdTrack}. Users may give away control over personal information as a result of the framing of an online service offer \cite{Angulo2014WWIT}. Additionally, there are multiple studies in the privacy decision-making literature that operationalize ``privacy control'' and ``perceived privacy control'' as either dependent or independent variables in their corresponding models (e.g., \cite{DinevXu2013IPC}, \cite{GriffinRajtmajer2016MPPBOU}, \cite{Krasnova2010}; see \cite{Shulman2018TBPDDMUCA} for a review of more papers on the topic).
		
		Technological implementations of control over privacy have been mostly concerned with cryptography and systems architecture (e.g., \cite{coppersmith1999}, \cite{Gong2016OTRMCPC}, \cite{jiang2002mpc}, \cite{Jones2010FSN}, \cite{sivaraman2015nlsecpricontrol} and others), or functionality enabling and interface design (e.g., \cite{Colnago2016IPA}, \cite{Knijnenburg2014IncreasingST}, \cite{Mehta2016PIS}, \cite{Schaub2015designspace}, \cite{Toch:2010:LPL}, and many more).\footnote{We invite our readers to explore independently the world of patents on privacy controls.}
		
		However, the privacy control literature so far has used the term ``control'' mostly as a mean of adjusting disclosure preferences (e.g., adjust settings, contact support, etc.), a level of disclosure adjustments that may be introduced (e.g., change settings and give or revoke consent in full or partially, on a level of a server, an application, a location, an enterprise, etc.), or as a plausible adjustment that can be made realistically (e.g., start disclosing, stop disclosing, delete information, etc.). Using a control theoretic analysis, and introducing a closed-loop control that can inform users about achieved disclosure outcomes or privacy states, we can start to use the term ``control'' more productively. With properly associated feedback we may know about achieved privacy states and disclosure outcomes. It may enable us to talk about ``controllability'' of privacy states and disclosure outcomes. Analysing controllability of privacy may help us answer questions about whether desired states or outcomes are reachable, and whether they have been reached.
		
		Empirical privacy research in computer science and human-computer interaction has already given some attention to feedback processes and their impact on privacy behavior and perceptions (e.g., \cite{Schlegel:2011:EYE} and \cite{Tsai2017TG} dealing with feedback design, \cite{Patil:2015:IIL} and \cite{Patil:2014:RAF} looking at effects of feedback recency).
		
		Trying to answer the question of how important a feedback mechanism can be for managing personal privacy, Tsai et al. \cite{Tsai:2009:WVY} demonstrate that the presence of feedback in a location-sharing scenario makes people feel more comfortable with disclosure of personal information and alleviates the level of privacy concerns. Thus, both the aforementioned increase in perceived control and the presence of feedback raise the people's information sharing propensity. These findings bear risks, alongside obvious benefits, and they should be treated with caution.
		
		In a paper concerned with the state of transparency enhancing technologies Murmann and Fischer-H{\"u}bner \cite{Murmann2017TAUEPT} provide a categorisation and assessment of existing transparency enhancing technologies, which greatly rely on feedback mechanisms. The authors note that without a feedback mechanism, users may be unable to make rational decisions about the use of transparency enhancing technologies and exercise control over them.
		
		Hoyle et al. \cite{Hoyle:2017:VVP} study relationships between content publishers and content users. Their findings lead to the conclusion that feedback mechanism may be a useful instrument for balancing personal information disclosure and exposure of the publishers' content. 
		
		Bargh et al. \cite{Bargh:2014:PPD} explore relationships between data controllers and data processors. The authors define ``feedback'' as any backwards-directed data flow from data processors to data controllers that facilitates forward-directed data flow. Their conceptual paper is concerned with the public policy discussion on procedural feedback between different agents dealing with personal data.
		
		Discussing nudges in privacy and security decision-making performed with the use of information, Acquisti et al. \cite{Acquisti:2017:NPS} distinguish between education and feedback, where education is responsible for affecting future decision-making, while feedback is capable of altering behavior at the current moment or over time. In terms of control theory, education corresponds to open-loop system dynamics, and feedback naturally relates to close-looped systems. It must be noted, however, that the authors use some colloquial understanding of the term ``feedback'', resulting in a debatable claim that ``feedback can also inform about expected and actual outcomes before or immediately after making a decision'' (\cite{Acquisti:2017:NPS}, p. 44:13). A process that informs about ``expected outcomes'' before an action perhaps constitutes a separate notion (e.g., predictive modelling, feed-forward control, predictive inference, hypothesising, etc. -- depending on the context), which is different from what is understood by feedback in control theory. 
		
		As we show, the term ``feedback'' in the privacy literature is often only loosely defined. If we want to proceed with analyses in the control theoretic framework, then we should align our understanding of the term ``feedback'' with the control theoretic definitions. One can use the following definition as an anchor point: \textit{Feedback} is ``the modification, adjustment, or control of a process or system (as a social situation or a biological mechanism) by a result or effect of the process, esp. by a difference between a desired and an actual result; information about the result of a process, experiment, etc.; a response'' \cite{OxfordEnglsihDictionary}. 
		
		Application of the control theoretic framework may not only imply the stricter definition of feedback. It is not any information related to privacy choices that is provided to the decision-maker. Control theoretic feedback returns information on achieved levels of outcomes, which decision-makers can compare to their own goal levels. This feedback hardly appears in a simple obvious way in reality. 
	
		As was mentioned before, the feedback mechanism can perhaps be implemented in technology. This technology, if it is built with control theoretic considerations, will differ from existing privacy-enhancing technologies and basic recommender systems. Existing systems provide recommendations derived from: 
		\begin{itemize}
			\item profiles of users and associations between profiles and specific users;
			\item the accumulated statistics (historical data) on privacy outcomes, which allow predictions of desired or unwanted outcomes for specific types of users; 
			\item the best practices and advice from scholars and professionals;
			\item the ``raw'' information about who, how and when can, may and will access the users' personal data, if they proceed with a given option;
			\item the same ``raw'' information about who, how and when (and possibly for what purpose) someone actually accessed specific users' personal data;
			\item and other external data.
		\end{itemize}
		
		The desired state of privacy, however, changes over time (a person may become more informed, more or less concerned, more or less alert, etc., as life changes). In order to resort to some comfort levels, a person needs to figure out what privacy-related action to perform. A person would need sufficient understanding of causal and temporal relationships between actions and privacy-entailing consequences, as well as of the character and form of these relationships. It is questionable that people are capable and willing to do that. Conversely, adjusting one's privacy to some comfort zone can be facilitated with the addition of a control-theoretic feedback loop, providing the following advantages:
		\begin{enumerate}
			\item Privacy outcomes of actions may be traced back to those actions in terms of cause and time, through the nature of feedback, accounting for time delays.
			\item Desired privacy outcomes may be compared with actual privacy outcomes.
			\item Effects of actions on privacy may be associated with these actions, even when the form of relationships between the actions and their effects is not proportional (more complex than one-to-one mapping, e.g., ``if-then'' rules). This is by accounting for physical lag and multiple elements and loops. 
		\end{enumerate}
		
		We face several issues, when we attempt to model the privacy feedback loop with control theory:
		\begin{itemize}
			\item Is there a way for the user or decision-maker to know and define their desirable outcomes and states of privacy?
			\item Is there a way for the user or decision-maker to associate feedback about privacy outcomes and implications with actions that have led to these outcomes and implications?
			\item Should users and decision-makers be nudged towards some optimal privacy configuration? What would be the optimality criteria in that case?
			\item Should users and decision-makers be nudged towards some specific privacy actions? What would be the justification for and against certain actions?
			\item General controllability issues highlighted in Table~\ref{tab1}. 
		\end{itemize}
		
		Technological implementation of the feedback loop may help people make better adjustments of their privacy behavior. The feedback may be partially approximated with quasi-linearity, modelled with anticipation (e.g., a quickened display), Kalman filter and finite state control with time lag and other concepts from control theory.
		
		Alternatively, we may model privacy as an open-loop system. The improvement of personal disclosure behavior may be achieved through enriching people's prior knowledge. One way to implement that is to provide relevant privacy education, and training.
		
		Thus, the control theory framework can be used to come up with an analysis of privacy and to inform the development of privacy solutions.
		
	\subsection{Future Work, Scope and Limitations} 
		The control theoretic approach provides various constructs and ideas to be tested in privacy-related research, including privacy attitudes and behaviors, especially decision-making.
				
		This conceptual analysis reveals several potential research directions:
		\begin{itemize}
			\item Study of feedback elements and their effects on privacy attitudes and behavior: time lag (delay) for feedback to flow between outcomes and actions, physical lag between an action and its effect on privacy, etc.
			\item Study of relations between different elements in a system, involving a user, the user's privacy state, the user's desirable privacy, information disclosure outcomes, evaluations of outcomes, external factors, and feedback loops between these elements.
			\item Study of users' decision-making, when feedback loops are involved.
			\item Modelling certain elements of the privacy decision-making process in more detail as separate control subsystems. Some of these subsystems may be suitable for more formal control theoretic representation (e.g., application or server permission management).
			\item Modelling individual differences in the control theoretic framework.
			\item Feedback-control loop implementation in practice.
			\item And, without a doubt, others.
		\end{itemize}
		
		We emphasize that, even though this paper is devoted to a conceptual control theoretic analysis of privacy, it can be naturally developed towards modelling and analyzing privacy control in information systems. Technological implementations of the privacy feedback loop, based on control theoretic principles, may facilitate individuals' control over personal information and its disclosure and may raise awareness of their current privacy states. 
		
		An applied control theoretic analysis of privacy may be appropriate and particularly valuable when it comes to the implementation of privacy-by-design principles. On the one hand, it may help in the evaluation of a system's compliance with the privacy-by-design principles via assessing controllability (and stability) of users' personal information disclosure. It may also consult the development of information systems with privacy-by-design in mind. On the other hand, an explicit feedback loop mechanism is easier to develop for a system, which is adhering to the privacy-by-design principles.
		
		We also note that this paper is not an exhaustive analysis of privacy in terms of control theory. We did not extend our paper with an alternative analysis, based on Powers' perceptual control theory, for the sake of keeping the scope and the rationale of the paper within reason and to avoid theoretical debates around PCT's assumptions and applicability in psychology. We also did not venture into the analysis of open-loop control of privacy. However, analyses of privacy in the literature so far have already treated privacy in a somewhat similar way to an open-loop system. It must also be noted that control theory is instrumental, when we deal with closed-loop control.
		
		The presented conceptual analysis of privacy as an object of control does not map the whole body of control theoretic constructs onto privacy research. We have omitted multiple domain- and application-specific concepts and tools. Our attempt has been to evaluate, transfer and adjust those control theoretic constructs that seem to bear benefits and can be fit to privacy-related research. For the sake of simplicity we also omitted more formal or specialized aspects and items (e.g., underlying partial differential equations, Kalman filter, feed-forward models, etc.), which still may be useful in further analyses of the subject and in relation to specific problems.      
				
	\section{Conclusion}
		
		In this paper we apply the theoretical framework of control theory to privacy, according to one of the major understandings of privacy as a person's control over information. We conceptualize privacy control with a human controller at its core, and raise questions about the controllability of such a system. 
		
		The conceptual model of privacy that we developed and presented in this paper allows us to reveal multiple controllability issues of privacy, and we propose several directions for future research with control theory in mind.
		
		We further discuss the relation and relevance of our proposed model and of the control theoretic analysis to privacy. We study the existing body of empirical privacy research and find multiple connections in how the privacy literature used and highlighted the notions of control and feedback. We also present our analysis of applicability of the proposed approach, as well as the challenges and opportunities of modelling personal information disclosure as a dynamic system with open and closed-loop control.
		
		One particular question we raise concerns the plausibility of a feedback control loop of privacy. If and when the implementation of a feedback control loop is infeasible, privacy may be analyzed as an open-loop control system. Our analysis shows that privacy may be a phenomenon that is inherently difficult to control. Some aids can perhaps be used to make it more controllable, such as indications about possible privacy implications of actions, or recommendations on privacy optimisation through the development of privacy-related feedback control loops. 
		
	\subsubsection{Funding.}
		This research is partially funded by the EU Horizon 2020 research and innovation programme under the Marie Sk{\l}odowska-Curie grant agreement No 675730 ``Privacy and Us''.
	  
	\bibliographystyle{splncs04}
	\bibliography{bibliography_ifipsc2018}
	
	\clearpage

\end{document}